\documentclass[twocolumn,prb,showpacs,amsmath,amssymb]{revtex4}
\pdfoutput=1
\usepackage{graphicx}
\usepackage{dcolumn}
\usepackage{bm}

\begin{document}

\title{Architecture for high-sensitivity single-shot readout and control of the electron
spin of individual donors in silicon}

\author{A. Morello$^{1,*}$, C. C. Escott$^1$, H. Huebl$^{1,\dag}$, L. H. Willems van Beveren$^1$,\\ L. C. L. Hollenberg$^2$, D. N. Jamieson$^2$, A. S. Dzurak$^1$, and R. G. Clark$^1$}

\affiliation{\\
$^1$ Australian Research Council Centre of Excellence for
Quantum Computer Technology, Schools of Electrical Engineering \& Telecommunications and Physics, University of New South Wales, Sydney NSW 2052, Australia,\\
$^2$ Australian Research Council Centre of Excellence for Quantum
Computer Technology, School of Physics, University of Melbourne,
Melbourne VIC 3010, Australia.}


\begin{abstract}
We describe a method to control and detect in single-shot the
electron spin state of an individual donor in silicon with greatly
enhanced sensitivity. A silicon-based Single-Electron Transistor
(SET) allows for spin-dependent tunneling of the donor electron
directly into the SET island during the read-out phase. Simulations
show that the charge transfer signals are typically $\Delta q
\gtrsim 0.2$e - over an order of magnitude larger than achievable
with metallic SETs on the SiO$_2$ surface. A complete spin-based
qubit structure is obtained by adding a local Electron Spin
Resonance line for coherent spin control. This architecture is
ideally suited to demonstrate and study the coherent properties of
donor electron spins, but can be expanded and integrated with
classical control electronics in the context of scale-up.
\end{abstract}

\pacs{73.23.Hk, 03.67.Lx, 71.55.Cn, 85.35.Gv, }

\maketitle

Electron spins bound to donor nuclei in silicon have exceptionally
long coherence and relaxation times, relative to the timescale for
the control of their quantum state,\cite{hill05PRB} and are thus a
promising qubit system. The electron spin coherence time of a
phosphorus donor is $T_{2} > 60$ ms at $T=6.9$~K in isotopically
pure $^{28}$Si,\cite{tyryshkin03PRB} already orders of magnitude
longer than for GaAs quantum dot systems\cite{petta05S,koppens06N}
and still far from the theoretical limit (dominated by the $^{29}$Si
impurity fraction,\cite{witzel05PRB} which can be minimized through
processing). However, a major obstacle to realizing a donor electron
spin qubit in silicon\cite{vrijen00PRA} is the difficulty of
detecting the spin state for individual donors typically 10-20 nm
below a SiO$_2$ interface.

The first proposals for donor spin readout involved spin-to-charge
conversion\cite{kane98N,hollenberg04PRB} through spin-dependent
tunneling to a doubly occupied D$^-$ donor state. The change in
electrostatic potential caused by an electron  can be detected by a
Single-Electron Transistor (SET) on the SiO$_2$ surface, as shown
e.g. in a double-donor well structure.\cite{andresen07NL} The
sensitivity of the detection scheme is quantified by the charge
transfer signal $\Delta q/e$, defined as the relative shift in the
SET bias point caused by the displacement of a nearby charge.
$\Delta q/e = 1$ if an electron is removed from the SET island and
taken to infinity. Clearly, $\Delta q/e < 1$ if the coupling between
donor and SET is purely electrostatic, i.e., no charge can be
directly transferred between the two. The exact value depends on how
far the electron can move to/from or in the vicinity of the SET
island: for a charge moving some 20 nm laterally in the silicon, 20
nm below the tip of the island of a surface SET, the signal is
typically very small, $\Delta q/e \sim 0.01$. This fact has
encouraged proposals in which $\Delta q/e$ can be increased by
confining the donor electron close to the SiO$_2$
interface,\cite{calderon06PRL} thus closer to the SET. On the other
hand, in quantum dot systems the successful readout of a single
electron spin has been achieved by monitoring the spin-dependent
tunneling of the electron into a reservoir.\cite{elzerman04N} Here
we take this concept a step further and present a donor-based,
electron spin qubit device, where the charge transfer signal upon
spin readout can be increased by over an order of magnitude as
compared to previous proposals. Our architecture combines recently
developed silicon quantum dot\cite{angus07NL} and
SET\cite{angus08APL} technologies with precise single-ion
implantation\cite{jamieson05APL} and local Electron Spin Resonance
(ESR).\cite{koppens06N,vanbeveren08APL} The crucial feature of this
device consists in using the island of a sub-surface silicon SET as
the electron reservoir for spin-dependent electron tunneling,
greatly enhancing the charge signal. We show that this method allows
fast (potentially in the nanosecond range) and high-sensitivity spin
readout with no back-action before the measurement, therefore
protecting the qubit from decoherence due to the measurement setup.

\begin{figure}[t] \center
\includegraphics[width=8.5cm]{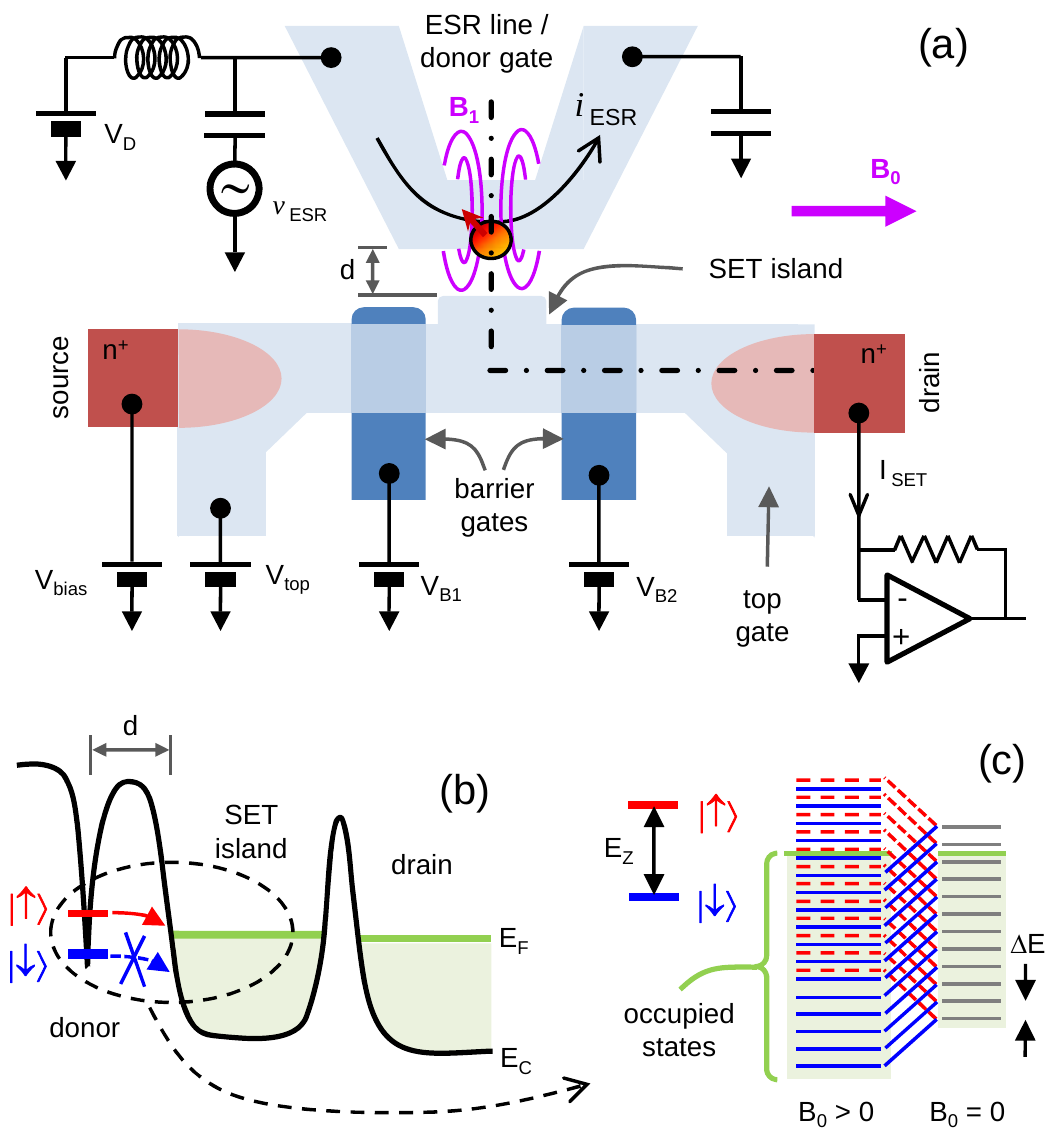}
\caption{ \label{Fig1} (Color online) (a) Schematic top view of the
spin qubit device. The donor-SET island distance is $d \sim 50$~nm.
(b) Energy profile along the dash-dotted line in panel (a). The
density of electron states in the SET island is approximated as a
quasi-continuum. (c) In reality, the SET has a finite number of
electrons: the quasi-continuum approximation and the spin readout
method are valid as long as $\Delta E \ll E_{\rm Z}$.}
\end{figure}

A sketch of the proposed donor spin qubit device is shown in
Fig.~1(a). It consists of three main elements: (i) a phosphorus
donor, introduced in the intrinsic Si substrate by single-ion
implantation;\cite{jamieson05APL} (ii) a
Si-SET,\cite{angus07NL,angus08APL} fabricated next to the donor
implant site, with a distance $d \sim 50$~nm between donor and SET
island; (iii) a coplanar transmission line, terminated by a short
circuit that runs just above the donor. The Si-SET comprises two
aluminum barrier gates and an overlaying top gate, insulated by
Al$_2$O$_3$ and deposited on high-quality SiO$_2$, typically 5 nm
thick. The top gate extends to source/drain n$^+$ doped regions, and
induces an electron layer under the SiO$_2$ when biased to $V_{\rm
top} \gtrsim 1$~V. Setting the barrier gate voltages to $V_{\rm
B1,2} < V_{\rm top}$, the bottom of the conduction band can be
lifted above the Fermi energy $E_{\rm F}$, thereby interrupting the
electron layer and forming the island of a SET [Fig.~1(b)].

The crucial difference between the Si-SET described here and the
more common surface Al-SETs is that the SET island now consists of a
small area of electron gas, electrostatically induced \emph{under
the SiO$_2$ layer}. Therefore, it is possible to construct a device
where \emph{an electron can tunnel between the donor and the island
of the SET}. This feature gives a charge transfer signal an order of
magnitude larger than achievable with other detection schemes, where
the charge sensor is only electrostatically coupled to the qubit. In
addition, the Si-SET has the advantage of being compatible with
standard Metal-Oxide-Semiconductor (MOS) fabrication processes,
since it does not require shadow metal evaporation.

The shorted coplanar transmission line serves a double purpose: (i)
it carries microwave pulses that produce an oscillating magnetic
field, $B_1$, for local ESR,\cite{koppens06N} and (ii) it acts as a
dc-gate for the electrostatic potential of the donor below it
through application of a static voltage $V_{\rm D}$. The ability to
use a transmission line to perform local ESR of donor spins while
simultaneously applying a dc-bias was recently demonstrated by
electrically-detected magnetic resonance at $T \ll
1$~K.\cite{vanbeveren08APL} Since the transmission line is
non-resonant, the structure is entirely broad-band. The purpose of
the short-circuit termination is to create a node of the electric
field at the donor site, while having the maximum value of the
current, $i_{\rm ESR}$, for fast coherent manipulation of the spin
state.\cite{koppens06N} If the short-circuit termination has a
cross-section $\sim 100 \times 100$~nm$^2$, a Rabi $\pi$-pulse can
be obtained in $< 100$~ns with less than -30 dBm microwave power on
the chip.

Figure~1(b) shows the energy profile along the dash-dotted line in
Fig.~1(a). First we encounter the potential well created by the
donor ion, which can bind one (D$^0$) or two (D$^-$) electrons. In
the following we shall restrict our analysis to the one-electron
D$^0$ state. By applying a static magnetic field $B_0$ in the plane
of the chip, the energy of the $|\uparrow\rangle$ and
$|\downarrow\rangle$ spin states is split by an amount $E_{\rm Z} =
2 g \mu_{\rm B} B_0 S$, with $g \approx 2$ and $S=1/2$. Proceeding
further we enter the SET island, where the bottom of the conduction
band $E_{\rm C}$ is pushed below the Fermi level $E_{\rm F}$ by the
positive $V_{\rm top}$. Between donor and SET is an energy barrier
that allows electron tunneling at a rate $\Gamma_{\rm D}$, mainly
determined by the distance $d$~($\sim 50$~nm) between donor and
island. Turning towards the drain contact, we cross the tunnel
barrier created by $V_{\rm B2} < V_{\rm top}$. This barrier is
easily tunable to have a tunnel rate $\Gamma_{\rm SET} \gg
\Gamma_{\rm D}$.

Spin-to-charge conversion is achieved by tuning $\{V_{\rm D},V_{\rm
top}\}$ so that only the energy of the $|\uparrow\rangle$ state is
above the electrochemical potential of a charge reservoir. The
detection of a charge transfer from donor to reservoir then
corresponds to the single-shot projective measurement of the
$|\uparrow\rangle$ state, as demonstrated in GaAs quantum
dots.\cite{elzerman04N} Here we propose to use the SET island as
electron reservoir, instead of having a separate bulk electron
layer. For this readout method to work, the Fermi distribution in
the SET island must be sharp on the energy scale set by $E_{\rm Z}$.
This condition is abundantly fulfilled by cooling the system to $T
\sim 100$~mK and using an operation frequency $\sim 40$~GHz for the
ESR system, corresponding to $B_0 \approx 1.4$~T and $E_{\rm
Z}/k_{\rm B} \approx 2$~K. In fact, due to the finite number of
electrons, the density of states in the SET island is not a real
continuum, but for the purpose of spin readout it is sufficient for
the single-particle energy spacing, $\Delta E$, to be much smaller
than $E_{\rm Z}$ and comparable to $k_{\rm B}T$ [Fig.~1(c)]. Taking
an SET island of area $A = 50 \times 100$ nm$^2$, we estimate
$\Delta E \sim 2 \pi \hbar^2 / g m^* A = 24$~$\mu$eV ($g=4$ is the
spin + valley degeneracy and $m^* = 8.9 \times 10^{-31}$~kg is the
effective mass), already smaller than the thermal broadening of the
Fermi function at $T \sim 100$~mK.

\begin{figure}[t] \center
\includegraphics[width=8.5cm]{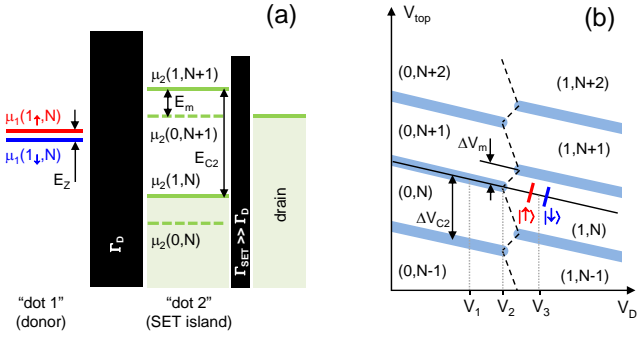}
\caption{ \label{Fig2} (Color online) (a). Sketch of the
electrochemical potentials in the ``parallel double dot'' picture.
The relevant SET potentials are the solid lines when the donor is
neutral (D$^0$ state), or the dashed lines when the donor is ionized
(D$^+$ state). (b) Charge stability diagram in the double dot
picture. The broad lines represent the SET current peaks, $I_{\rm
SET} > 0$, spaced by $\Delta V_{C2} = E_{\rm C2} / e \alpha_2$ along
the $V_{\rm top}$-axis, where $\alpha_2$ is the lever arm between
top gate and SET island. The dashes labeled
$|\uparrow\rangle,|\downarrow\rangle$ represent the positions of
$\mu_1(1_{\uparrow},N)$ and $\mu_1(1_{\downarrow},N)$ as given in
panel (a). The $\mu_2$ ladder can be kept fixed by moving $\{V_{\rm
D},V_{\rm top}\}$ along the thin black line.}
\end{figure}

The advantage of having an integrated charge sensor and electron
reservoir is best appreciated by thinking of this architecture as a
``parallel double quantum dot'' configuration.\cite{hofmann95PRB} We
label the donor as ``dot 1'', which can only have 0 or 1 electrons,
and the SET island as ``dot 2'', with a large number of electrons
$N, N+1, \ldots$ and charging energy $E_{\rm C2} \sim
1$~meV.\cite{angus08APL} The charging energy of the donor is $E_{\rm
C1} \sim 30 -  40$~meV, depending on the capacitance to its
surroundings.\cite{lansbergen08NP} Dot 2 is connected to source and
drain contacts and can be measured in transport, while the charge
state of dot 1 can only be changed by electron tunneling to or from
dot 2.

For two quantum dots with mutual coupling energy $E_{\rm m}$, the
electrochemical potential of one dot depends on the charge state of
the other.\cite{vanderwiel03RMP} Therefore the SET can have two
ladders of electrochemical potentials, shifted by $E_{\rm m}$ with
respect to each other, depending on the charge state of the donor.
The donor states are split by $E_{\rm Z} = \mu_1(1_{\uparrow},N) -
\mu_1(1_{\downarrow},N)$. In Fig.~2(a) we sketch a situation where
the charge state of the ``double dot'' is $(1,N)$ (the donor is
neutral), thus the relevant electrochemical potentials on the SET
side are the solid lines, $\mu_2(1,N), \mu_2(1,N+1), \ldots$ and the
corresponding $\{V_{\rm D},V_{\rm top}\}$ point in Fig.~2(b) is
given by the dashes labeled $|\downarrow\rangle,|\uparrow\rangle$.
The SET is thus in Coulomb blockade, with $I_{\rm SET}=0$, because
$\mu_2(1,N+1)$ is far above the Fermi level in the source/drain
contacts.

Next we wish to lift $\mu_1$ by decreasing $V_{\rm D}$, while
keeping the $\mu_2$ ladder fixed and $\mu_2(0,N+1)$ in the
source/drain bias window. Because of cross-capacitance, this
requires a compensating $V_{\rm top}$ to keep the operation point
$\{V_{\rm D},V_{\rm top}\}$ along the solid line in Fig.~2(b). When
$\mu_1(1,N) > \mu_2(0,N+1)$, the donor electron can tunnel onto the
SET island (at a rate $\Gamma_{\rm D}$) and out to the drain (at a
rate $\Gamma_{\rm SET} \gg \Gamma_{\rm D}$), leaving behind a
positive charge at the donor site. This charge ``pulls down'' the
ladder of electrochemical potentials for the SET, now represented by
the dashed lines in Fig.~2(a), $\mu_2(0,N), \mu_2(0,N+1), \ldots$
and since $\mu_2(0,N+1)$ is in the source-drain bias window, $I_{\rm
SET}$ jumps to its maximum value. The charge transfer signal is
$\Delta q / e = \Delta V_{\rm m} / \Delta V_{\rm C2}$ [Fig.~2(b)],
i.e. the shift of the SET Coulomb peaks due to the ionization of the
donor, divided by their period. Note that $\Delta q / e$ is exactly
what would be obtained by moving a positive charge from infinity to
the donor site. Thus, $\Delta q / e \rightarrow 1$ as the donor
location approaches the SET island.

\begin{figure}[t] \center
\includegraphics[width=8.5cm]{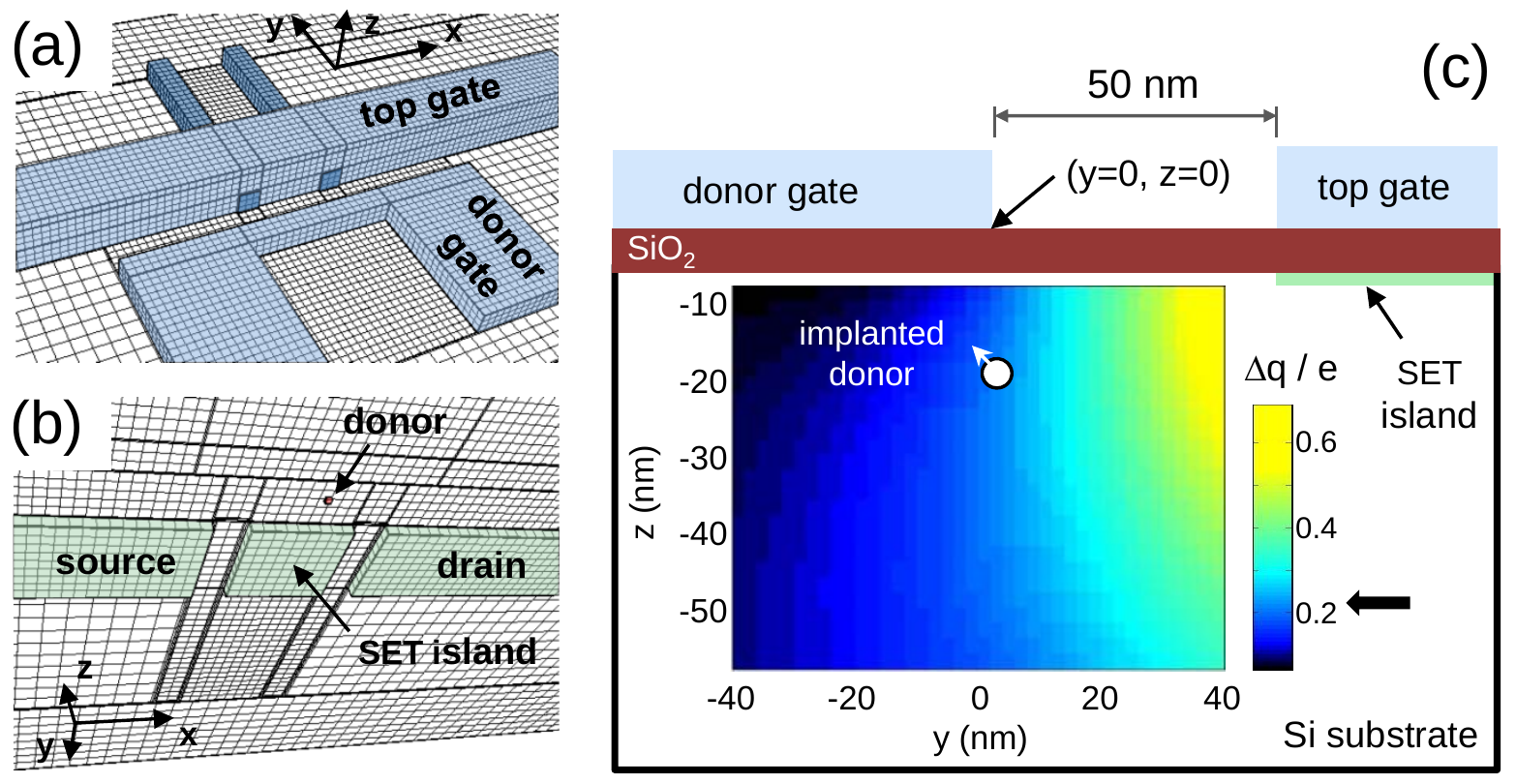}
\caption{ \label{Fig3} (Color online) Top (a) and bottom (b) view of
the device model used to calculate the charge transfer signal
$\Delta q / e$. Filled areas represent gates and electron layers as
indicated. (c) FastCap calculation of $\Delta q / e$ as a function
of donor position in the $y-z$~plane. We used a mesh 5 times finer
than shown by the thin lines in (a) and (b), yielding discretization
errors $< 5$~\% of the $\Delta q / e$ value at each point. For a
donor under the tip of the donor gate, 15 nm below the Si/SiO$_2$
interface, $\Delta q / e \approx 0.2$ (black arrow).}
\end{figure}

We have set up a device model, shown in Fig.~3(a,b), for use in the
boundary-element, capacitance extraction code
FastCap\cite{nabors91IEEE} to determine coupling capacitances as a
function of donor position. In the model the gates and electron
layers are described by metallic conductors of the appropriate size
and the donor is represented as a metal cube with sides of length
$2a_{\rm B}$ ($a_{\rm B} \approx 2.5$~nm is the Bohr radius in
silicon). $\Delta q / e$ can be expressed as the ratio $C_{\rm m} /
C_{1\Sigma}$, where $C_{\rm m}$ is the mutual capacitance between
donor and SET island, and $C_{1\Sigma}$ is the total donor
capacitance. Figure~3(c) shows the resulting $\Delta q / e$,
assuming 50~nm gap between top gate and ESR line, 5~nm SiO$_2$
thickness, and donor in the $y-z$~plane. For a donor right under the
tip of the ESR line, and 15 nm below the Si/SiO$_2$ interface, we
find $\Delta q / e \sim 0.2$. With $E_{\rm C2} \sim 1$~meV and $T
\sim 100$~mK, $I_{\rm SET}$ can shift from zero to its maximum
value, $I_{\rm max}$.

A typical spin control and readout sequence would proceed as shown
in Fig.~4, always assuming the $\mu_2$ ladder is kept fixed by using
compensated $\{V_{\rm D},V_{\rm top}\}$ pulses. (i) Empty: the donor
is ionized when $V_{\rm D} = V_1$, causing $\mu_1(1_{\downarrow},N)
> \mu_2(0,N+1)$. The successful donor ionization is signalled by
$I_{\rm SET} = I_{\rm max}$. (ii) Load: a new electron is loaded
into the ground Zeeman state $|\downarrow\rangle$, by choosing $V_2$
such that $\mu_1(1_{\uparrow},N) > \mu_2(0,N+1) >
\mu_1(1_{\downarrow},N)$, and $I_{\rm SET} = 0$. (iii) ESR pulse:
when $V_{\rm D} = V_3$ both donor levels are far below
$\mu_2(0,N+1)$. The spin undergoes coherent Rabi rotations under the
effect of microwave pulses applied to the ESR line. Here we take the
example of a $\pi$-pulse where the final state is
$|\uparrow\rangle$. (iv) Readout: $V_{\rm D} = V_2$, and since
$\mu_1(1_{\uparrow},N) > \mu_2(0,N+1)$, the electron in the
$|\uparrow\rangle$ state tunnels off the donor into the SET island,
unblocking the conduction. However, since $\mu_2(0,N+1) >
\mu_1(1_{\downarrow},N)$, another electron can tunnel onto the donor
in the state $|\downarrow\rangle$, blocking the SET again. Thus, an
electron in $|\uparrow\rangle$ is signalled by a ``blip'' in $I_{\rm
SET}$, with a duration of order $\Gamma_{\rm D}^{-1}$. The
inhomogeneous spin coherence time, $T_2^*$, can be extracted by
observing the decay of Rabi oscillations, obtained by counting the
occurrence of $|\uparrow\rangle$ states as a function of ESR pulse
duration. The spin-lattice relaxation time, $T_1$, can be obtained
by loading an electron in an unknown state, i.e., using $V_{\rm D} =
V_3$ for the load pulse, then counting the occurrence of
$|\uparrow\rangle$ states as a function of the waiting time between
load and measurement\cite{elzerman04N} (no ESR required).

An important feature of our device is that, due to Coulomb blockade,
$I_{\rm SET}=0$ for the whole time the electron resides on the
donor, i.e., the charge sensor is automatically switched off. The
type of back-action arising, e.g., from the current through a
quantum point contact used as charge sensor,\cite{gustavsson07PRL}
is therefore eliminated.

\begin{figure}[t] \center
\includegraphics[width=8.5cm]{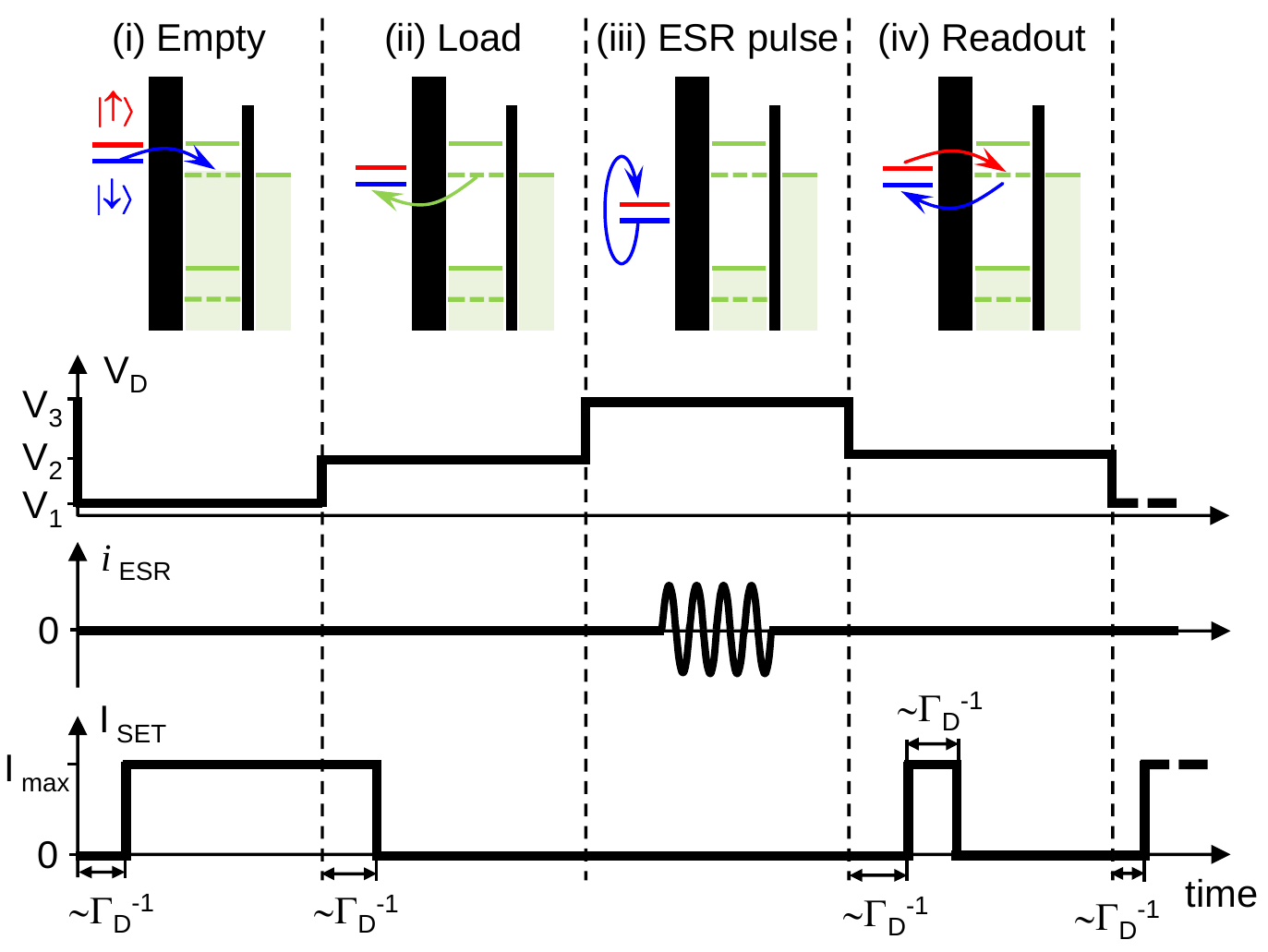}
\caption{ \label{Fig4} (Color online) Pulsing protocol and
corresponding SET signal for the coherent control and readout of a
donor electron spin. The donor gate voltages $V_{1,2,3}$ are shown
also in Fig.~2(b).}
\end{figure}

The timescale for projective readout of the qubit is set by the
typical tunneling time between donor and SET island, $\Gamma_{\rm
D}^{-1}$. In the design discussed here, $\Gamma_{\rm D}$ is
exponentially sensitive to $d$, but some level of control could be
achieved by introducing an additional gate between ESR line and SET.
However we note that the acceptable range of $\Gamma_{\rm D}$ for
reliable spin readout is extremely wide. The upper bound to
$\Gamma_{\rm D}^{-1}$ is set by the spin-lattice relaxation time,
$T_1$, since an electron in the $|\uparrow\rangle$ state must be
measured before it decays to $|\downarrow\rangle$. The bulk value
for P in Si is $T_1 \approx 3000$~s at $T = 1.25$~K, further
increasing as $1/T$.\cite{feher59PR} For a donor placed near a
Si/SiO$_2$ interface, paramagnetic defects and charge traps may give
an additional contribution to the spin relaxation and
dephasing.\cite{schenkel06APL} However, this contribution vanishes
exponentially when $k_{\rm B}T \ll E_{\rm Z}$ due to high spin
polarization.\cite{desousa07PRB} The lower bound to $\Gamma_{\rm
D}^{-1}$ is simply set by the bandwidth of the circuit that detects
$I_{\rm SET}$. Here we note that an ideal readout scheme would
exploit the fact that our system effectively provides a digital
signal ($I_{\rm SET} = 0$ or $I_{\rm max}$) by connecting the SET to
a cryogenic current comparator.\cite{gurrieri08IEEE} This method
could yield the ultimate readout speed, potentially as fast as $\sim
1$~ns. The MOS-compatibility of our design is significant, because
it allows the on-chip integration of qubit devices and ultra-fast
cryogenic readout electronics, using the same industry-standard
fabrication process. This is extremely advantageous both for readout
speed - minimizing the capacitance of the interconnects - and for
the scale-up of a quantum computer.

In conclusion, we have shown that recently-developed Si-SET
technology\cite{angus07NL,angus08APL} can be harnessed to create a
compact donor spin qubit architecture, wherein the island of the SET
is used as a reservoir for high-sensitivity spin readout. The
compatibility with MOS fabrication techniques is very advantageous
to integrate the qubits with digital on-chip
electronics\cite{gurrieri08IEEE} and achieve the ultimate in readout
speed and sensitivity for spins in the solid-state. This
architecture removes a major impediment to exploiting the natural
advantages of donors in silicon as a platform for scalable qubit
science.

\begin{acknowledgments}
This work was funded by the Australian Research Council, the
Australian Government, the U.S. National Security Agency, and the
U.S. Army Research Office under Contract No. W911NF-04-1-0290.\\
$^*$ Electronic address: a.morello@unsw.edu.au\\
\dag Present address: Walther-Meissner-Institut, Bayerische Akademie
der Wissenschaften, 85748 Garching, Germany.

\end{acknowledgments}


\end{document}